\documentclass[preprint,12pt]{elsarticle}

\usepackage{amssymb}
\usepackage{multirow}
\usepackage{booktabs} 
\usepackage{url} 
\usepackage{amsmath}
\usepackage{subcaption}

\journal{Visual Informatics}

\begin{document}

\begin{frontmatter}



\title{Metabook: A Mobile-to-Headset Pipeline for 3D Story Book Creation in Augmented Reality}


\author[hkustgz]{WANG Yibo\corref{cor1}}
\ead{ywang026@connect.hkust-gz.edu.cn}

\author[hkustgz]{MAO Yuanyuan}
\ead{ymao709@connect.hkust-gz.edu.cn}

\author[polyu]{Lik-Hang Lee}
\ead{lik-hang.lee@polyu.edu.hk}

\author[hkustgz]{NI Shi-ting}
\ead{stni289@connect.hkust-gz.edu.cn}

\author[hkustgz]{WANG Zeyu}
\ead{zeyuwang@ust.hk}

\author[hit]{GU Xiaole}
\ead{guxl@hit.edu.cn}

\author[hkustgz]{HUI Pan}
\ead{panhui@ust.hk}

\cortext[cor1]{Corresponding author}

\affiliation[hkustgz]
{organization={The Hong Kong University of Science and Technology(Guangzhou)},
            addressline={ No.1 Du Xue Rd, Nansha District}, 
            city={Guangzhou},
            postcode={511453}, 
            state={Guangdong},
            country={China}}
\affiliation[polyu]{
  organization={Hong Kong Polytechnic University},
  city={Hong Kong},
  country={China}
}
\affiliation[hit]{
  organization={Harbin Institute of Technology},
  city={Harbin},
  country={China}
}

\begin{abstract}
AR 3D book has shown significant potential in enhancing students' learning outcomes. However, the creation process of 3D books requires a significant investment of time, effort, and specialized skills. Thus, in this paper, we first conduct a three-day workshop investigating how AI can support the automated creation of 3D books. Informed by the design insights derived from the workshop, we developed Metabook, a system that enables even novice users to create 3D books from text automatically. To our knowledge, Metabook is the first system to offer end-to-end 3D book generation. A follow-up study with adult users indicates that Metabook enables inexperienced users to create 3D books, achieving reduced efforts and shortened preparation time. We subsequently recruited 22 children to examine the effects of AR 3D books on children's learning compared with paper-based books. The findings indicate that 3D books significantly enhance children's interest, improve memory retention, and reduce cognitive load, though no significant improvement was observed in comprehension. We conclude by discussing strategies for more effectively leveraging 3D books to support children's learning, and offer practical recommendations for educators.
\end{abstract}



\end{frontmatter}



\section{INTRODUCTION}
\label{sec1}
\begin{figure}[htbp]
\includegraphics[width=\linewidth]{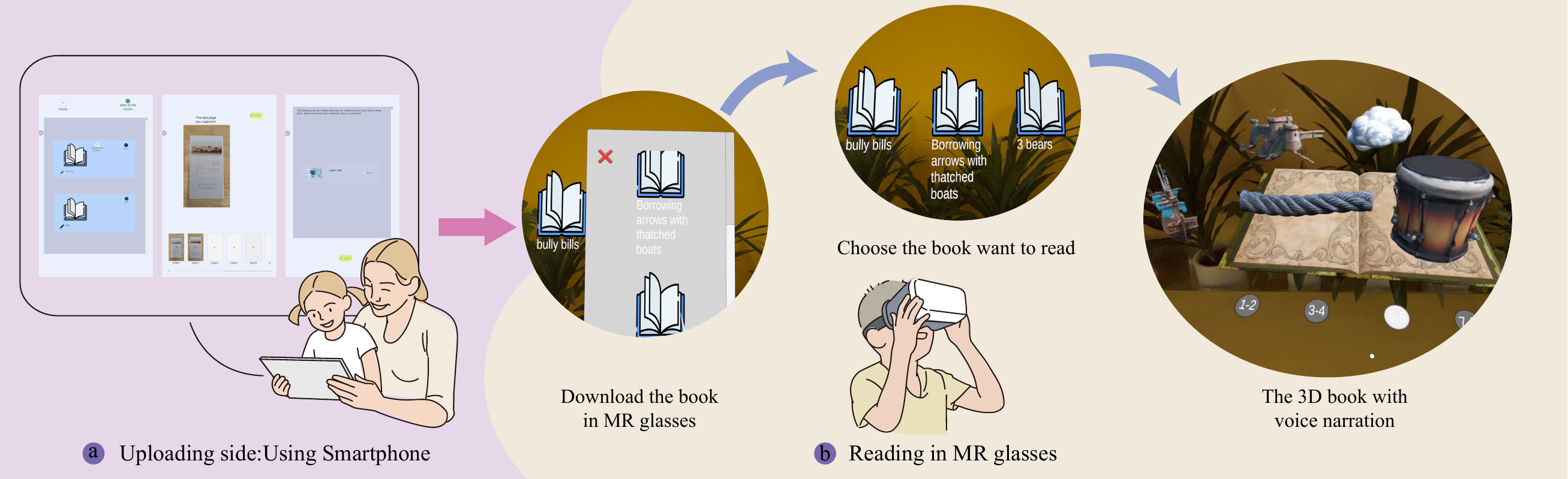}
  \caption{An Overview of Metabook, a system that enables the end-to-end creation of AR 3D books from text. (a) Users upload books
  in PDF or image format using their smartphones. (b) Users enjoy an immersive experience of the 3D version of the book.}
  \label{fig:teaser}
  \end{figure}
AR 3D book plays a pivotal role in supporting children's cognitive development and early learning by providing authentic, situational, and multisensory storytelling experiences. By combining interaction with spatially organized visual narratives, 3D books can enhance engagement and attention through a combination of audio and visual elements~\cite{dias2009technology}.  However, the manual creation process of 3D books requires a significant investment of time, effort, and specialized skills, which has hindered AR 3D books from being widely adopted. 
To simplify the creation of 3D books, previous work has proposed providing templates for 3D book creation \cite{10.1145/1690388.1690484}. However, this approach requires users to source, create, or edit models one by one. The advent of long-text-based 3D generation offers promising potential to significantly simplify the creation process of 3D books \cite{ocal2024sceneteller,zhou2024gala3d,huang2024toward}. However, these approaches are limited to generating 3D content from long textual descriptions of scenes and are unable to effectively process reading materials, which often contain complex elements such as contextual narratives, historical backgrounds, and specific character details.  As a result, these approaches are not well-suited for generating 3D books directly from story texts.

To this end, we propose a research question \textbf{(RQ1): How can we create 3D books directly from story texts?}
To explore this question, we conducted a three-day workshop with domain experts to explore the AR 3D book creation in the era of AIGC. During the workshop, we investigate: A) a way of leveraging AI models to replicate the human process of creating an AR 3D book; B) a method of achieving high consistency of content between the generated AR book and the story content and context; C) an approach to ensure the plausibility of the generated models. The investigation above led to four design guidelines (D1 -- D4). Accordingly, we designed and implemented Metabook, a system that enables novice users to create AR 3D books from story text. In the automatic pipeline of Metabook (Figure \ref{fig:teaser}), users can create their 3D book in three key stages. First, a user can upload the reading material via smartphone. Second, the user can review suspicious content, i.e., problematic 3D models flagged by an AI detection. Finally, the user can download the generated 3D books into the mixed reality (MR) glasses. To our knowledge, Metabook is the first system that enables the end-to-end creation of 3D books.
To evaluate the usefulness of Metabook's book generation, we conducted a user study in which eleven participants were asked to create their AR 3D books. The results indicate that Metabook makes it easy for users without AR 3D book creation experience to create AR 3D books, in addition to the benefits of reducing manual effort and time investment. In other words, users are able to produce satisfying AR 3D books within a satisfactory time window. 

Furthermore, we investigated the educational effectiveness of the generated 3D books. As such, our second research question \textbf{(RQ2) is: How does AR 3D book affect children's learning compared to paper books?}
We conducted a five-day within-subject comparative user study with 22 children, where participants experienced both the generated 3D book and the traditional paper book. We measured their interest, retention of the story, comprehension of the story, and cognitive load under both conditions. In addition, we conducted semi-structured interviews with six teachers to explore the potential value of our system in educational settings.
By combining objective performance metrics with subjective user feedback, our research offers practical insights for educators on how to leverage AR 3D books better to enhance children's learning outcomes. Metabook also contributes to the development of more accessible and intelligent authoring tools that integrate AR and AI, aiming to lower the barriers for non-experts to create immersive AR content. 
In summary, our contributions are as follows:
\begin{itemize}
    \item We conducted a three-day workshop with experts (N = 6) to explore  how AI can support the end-to-end creation of 3D books, an area that remains underexplored. The findings provide valuable insights for designers and researchers developing  authoring tools that integrate AR and AI in the future.
    \item Informed by the workshop, we designed and implemented an interactive prototype, Metabook. Metabook allows users to upload reading materials via smartphone to  create 3D books from text, which can then be downloaded and experienced in MR glasses. Our user study (N = 11) demonstrates that Metabook makes it easy for adult users without 3D book creation experience to create 3D books, while reducing both manual effort and time investment.
    \item We conducted a comparative user study with children (N = 22) to comprehensively investigate the impact of using 3D books on children's learning. We also carried out semi-structured interviews with teachers. Based on the results of the study and interviews, we discussed the practicality and usability of the generated AR 3D books and provided design recommendations for educators and developers of AR education system.
\end{itemize}
\section{Related Work}
\textbf{Text to 3D Generation. }
Advancements in machine learning, particularly generative models like GANs~\cite{goodfellow2020generative} and VAEs~\cite{kingma2013auto}, have opened new avenues for 3D generation. Projects like Pix2Vox~\cite{xie2019pix2vox}, PointNet~\cite{qi2017pointnet}, and Neural Radiance Fields (NeRF)~\cite{mildenhall2021nerf} demonstrate the potential of neural networks to generate 3D shapes and scenes from 2D images or sparse data. While promising, these methods are still limited in terms of resolution, realism, and consistency.
To overcome the scarcity of 3D training data, recent research has explored optimization-based generation~\cite{li2024advances}.  One of the representative works is DreamFusion~\cite{poole2022dreamfusion}. The 3D models generated by these works are of good quality, but slow to generate, so it is difficult to apply them to human AR interaction systems directly.

Real-time 3D generation is becoming increasingly relevant for applications like gaming, and MR. Feed-forward 3D reconstruction like TripoSR~\cite{tochilkin2024triposr} facilitates rapid 3D model generation through fast feed-forward inference~\cite{groueix2018papier,huang2023shapeclipper,huang2023zeroshape,li2023instant3d,tang2024lgm,wang2018pixel2mesh,wang2023pf}. Introducing TripoSR for 3D generation in MR allows for the rapid creation of MR scenes, thereby ensuring a satisfying user experience.  Although these real-time generations are rapid, they cannot handle long texts like those found in a storybook. Recent works \cite{ocal2024sceneteller,zhou2024gala3d,huang2024toward,chang2017sceneseer} explored how to use large language models to process long texts to generate 3D scenes that include more models. However, these works are limited to long descriptive texts and require positional relationships between models to be specified in the description. However, reading materials, especially stories, are often narrative in nature, with the portrayal of characters and objects typically embedded in the broader context. Moreover, stories often carry significant historical context. Therefore, these works are less suitable for generating models that align with the context of reading content. Our user-centric system, through step-by-step prompt engineering, CLIP-based verification, and manual confirmation of suspicious models, makes real-time book generation applicable to long texts in reading materials.

\textbf{AR 3D Books.} 
Augmented Reality (AR) possesses numerous advantages over traditional media, establishing it as a powerful tool for both educational assistance and storytelling.  As early as 1998, research on AR books emerged \cite{rekimoto1998matrix}. Currently, studies on AR book systems encompass a wide range of fields, including history \cite{15}, biology \cite{8}, electromagnetics \cite{11}, spatial ability development \cite{29}, astronomy \cite{36}, mathematics \cite{mathsAR}, and Chinese poetry \cite{poetryAR}. AR books provide realistic, interactive environments that simulate authentic communication scenarios. With the development of mobile technology, AR books enable students to immerse themselves in situational learning anytime and anywhere. The AR-Poetry system~\cite{arpoetry}, for example, focusing on classical Chinese landscape poetry, allows children to explore the landscapes described in the poetry from different perspectives in 3D books. Regarding storytelling, AR also serves as a medium to enhance the immersive experience of storybooks, providing a visual experience that surpasses paper-based books~\cite{magicbook,33,37}. 
Although many prior studies have shown the advantages of 3D books in AR (i.e., 3D AR Books), AR 3D books have failed to gain widespread use. The primary obstacle is the tedious creation process of AR 3D books that require a significant investment of time, effort, and specialized skills, which has hindered AR 3D books from being widely adopted. Artutor enables users without programming experience to edit AR books. However, it relies on existing model resources. An AR-Poetry system \cite{arpoetry} simplifies AR 3D book creation by using text recognition to match the recognized text with 3D models sourced from a database and display them. However, since the models are sourced from databases, if the database lacks suitable models, the AR book creation will fail. This greatly limits the scope of AR books that can be produced.   Our work bridges this gap by proposing an end-to-end story for the 3D book pipeline.

\textbf{The Impact of AR 3D Books on Children's Leading. }
AR 3D books can influence children's learning in multiple ways. In terms of comprehension, studies have shown that AR 3D books may reduce the difficulty of recognizing unfamiliar words \cite{danaei2019influence,danaei2020comparing}, thereby enhancing overall understanding of the text. For example, Delneshin Danaei et al. \cite{danaei2020comparing} found that AR 3D books can enhance children's comprehension, including improving their understanding of story structure and increasing their ability to answer implicit questions. Regarding cognitive load, some works argue that AR 3D book reading can be cognitively demanding \cite{chang2024embedding,radu2014augmented,munzer2019differences,10674259}. The sound, images, and various functions may serve as distractors for young readers \cite{radu2014augmented}.  However, under such contradictory views, the latest research indicates that AR books can alleviate children's cognitive load by boosting their confidence and providing a better flow experience \cite{cheng2017reading,wu2024effects}. The reason for this difference may lie in the design of the 3D books. Some 3D books have poorly designed features and functions, making them difficult for children to use, which may increase cognitive load \cite{10674259,csimcsek2024examining}. In terms of recalling a story, previous studies have shown that AR 3D books can enhance children's memory of stories, leading to better performance when they retell the story \cite{danaei2020comparing,du2024impact}. AR 3D books can also increase children's interest through multimedia elements such as animations, interactions, and sounds \cite{rambli2013fun,dibrova2016ar,chemerys2022combined,lukcyhasnita2024development}.
The existing work usually focuses on separate learning outcomes, such as children's comprehension, memory, and interest. In contrast, our work provides a comprehensive evaluation of comprehension, memory, interest, and the user's cognitive load. Notably, our work reinforces our understanding of AR 3D books with the educational outcomes and user perspective in the era of generative AI. 
\section{Workshop: Collecting Design Requirements} \label{sec:workshop}
We conducted a three-day workshop with six participants (denoted as I1 -- I6) to explore  how AI can support the automated creation of 3D books. The workshop's primary goal is to enhance our understanding of the process of creating 3D books from stories. As such, the participants, with either illustration design or generative AI background, were recruited through social media.  Among all participants, I1 to I5 are professional illustrators with more than three years of experience. I6 is a third-year PhD student whose research focuses on generative AI. 
Throughout the workshop, we focused on three core themes, and each day of the workshop ran a theme. On the first day, we investigated the use of AI models (e.g., LLM, 3D model generation) to replicate the human process of producing an AR book. On the second day, we examined methods to enhance the alignment of the created models with the context and narrative. On the third day, we analyzed methods to ensure the reliability of the generated 3D models. After the workshop, the first author reviewed the workshop recordings and conducted open coding, followed by a discussion with the senior author, during which they ultimately reached a consensus.
\subsection{Leveraging AI for replicating book creation}
We use semi-structured interviews with I1 to I5 to investigate the essential components to be included in drawings and the manner in which these components should be represented. Ultimately, we proposed a set of design principles to guide the transformation of narrative texts into 3D models embedded in 3D books.
\paragraph{Procedure.} Initially, we posed the following two enquiries to the participants: What components do you often use while producing illustrations? From which perspectives do you illustrate these elements?  Subsequently, two researchers performed open coding on the interview tapes and achieved the following results.  Based on the results, we also requested I1--I5 to analyze five stories. Each individual should identify the principal characters and key things from their designated narratives. By the conclusion of the day, we had collected 25 examples from them that were applicable to prompt engineering.
\paragraph{Results.} Main characters and main objects from the story book should be included.  All of them mentioned the importance of the main characters and would handle them separately. I4: ``Unlike animation, where there are many frames per second, illustration leaves out many scenes for the reader to imagine. As long as the character is well-drawn, readers can easily imagine the scenes of the story while reading. That’s why I always design the characters first.''  Regarding drawing the scene, we found that everyone mentioned only illustrating the main objects in the scene. I5: ``Due to publication space limitations, we don't depict all the objects in a scene. When there's more space, we include more objects; when space is limited, we focus on illustrating only the most essential ones.'' The main objects they chose when illustrating a scene varied based on their personal artistic style. I1: ``I usually draw the objects that the main characters are currently using.'' I3: ``I tend to focus more on the environment surrounding the main characters, such as depicting a forest or a lake.'' 

They shape characters through gender, nationality, age, recognizable appearance features, clothing, and the era of life. 
I2 gave an example: ``If I just draw a boy with green eyes, it’s hard to imagine he’s Harry Potter. But if I give him round glasses and a lightning-shaped scar on his forehead, anyone who has read the book will immediately recognize him.''

\paragraph{Design Requirements.}
Based on the results from the first-day workshop, we summarize design guidelines \textbf{D1}: We need to extract the main characters and main objects from the story;
and, \textbf{D2}: The main characters need to be depicted based on gender, nationality, age, recognizable appearance features, clothing, and the era of life.

\subsection{Alignment between Context \& Story in 3D Generation}
We conducted a comparative experiment to evaluate how the inference capabilities of LLMs can assist 3D-model-generation models in producing contextually appropriate 3D models.
\paragraph{Apparatus.} We used GPT-4 as the LLM and TripoAI \cite{tochilkin2024triposrfast3dobject} for 3D model generation. Both models were accessed via API calls. Our test set was derived from the 25 examples provided by I1--I5 during the first-day workshop. 

\begin{figure}[htbp]
    \centering
    \begin{subfigure}[t]{0.23\textwidth}
        \centering
        \includegraphics[width=\linewidth]{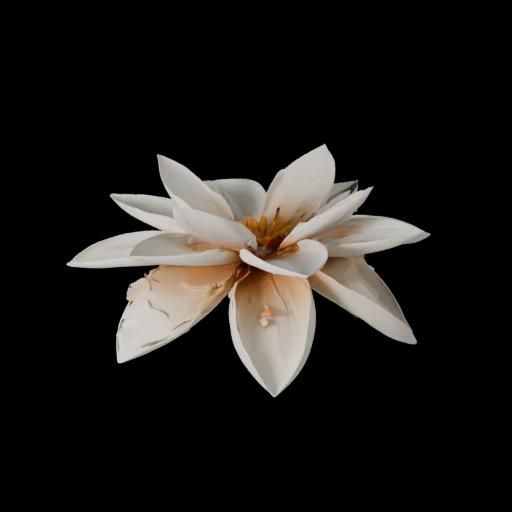}
        \caption*{(a)``golden white flower'' in \textit{Narcissus and the Narcissus Flower}, without GPT-4 Explanation}
    \end{subfigure}
    \begin{subfigure}[t]{0.23\textwidth}
        \centering
        \includegraphics[width=\linewidth]{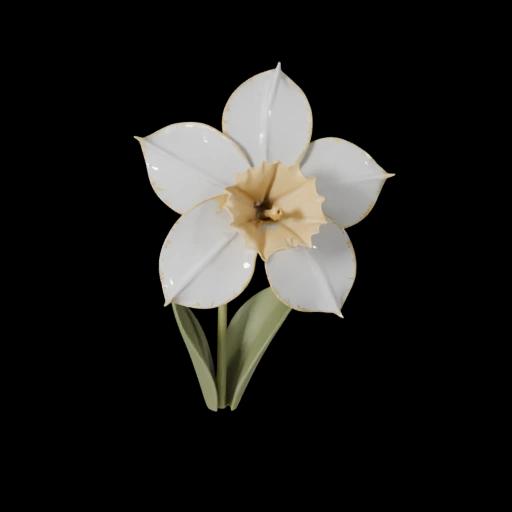}
        \caption*{(b)``golden white flower'' in \textit{Narcissus and the Narcissus Flower}, with GPT-4 Explanation}
    \end{subfigure}
    \begin{subfigure}[t]{0.23\textwidth}
        \centering
        \includegraphics[width=\linewidth]{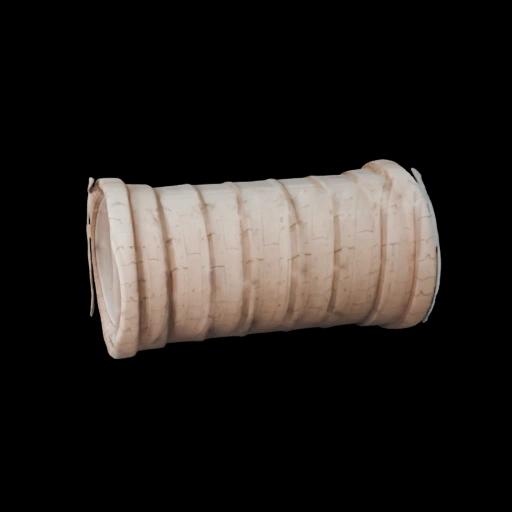}
        \caption*{(c)``tuba'' in \textit{learning to say no}, without GPT-4 Explanation}
    \end{subfigure}
    \begin{subfigure}[t]{0.23\textwidth}
        \centering
        \includegraphics[width=\linewidth]{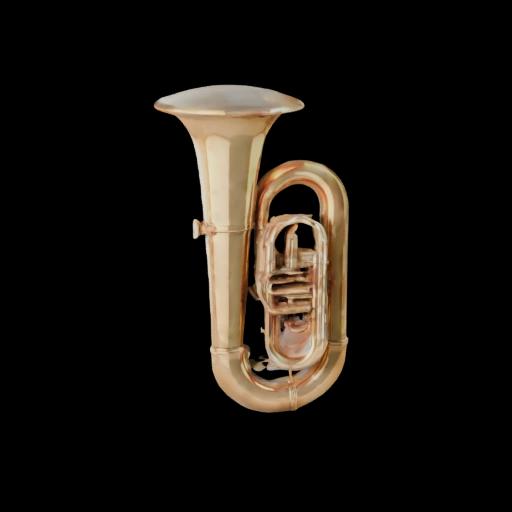}
        \caption*{(d)``tuba'' in \textit{learning to say no}, with GPT-4 Explanation}
    \end{subfigure}
    \caption{The comparisons of the 3D models.}
    \label{workshop}
    \vspace{-1mm}
\end{figure}

\paragraph{Procedure.}
We first conducted a baseline assessment in the initial round. We asked I6 to generate the 3D models of extracted characters and main objects. Then, I1 -- I5 rated the generated models on a scale from 1 to 5, where 1 indicates a poor level of matchness with the story and 5 indicates a very strong level of matchness. Next, we evaluated the experimental group. I6 were told to use GPT-4 to describe the main characters and main objects.  The prompt should include the full story, an explanation and a description of the object. Then I6 sent the main objects, along with their corresponding descriptions, to the TripoAI API for the second round of generation.  Afterwards, I1 – I5 rated the generated models following the aforementioned 5-point scale. 
\paragraph{Results.} In the baseline and experimental groups, the generated models received scores of 3.2 out of 5 and 4.6 out of 5, respectively. This indicates that providing descriptions of the objects before generation can help produce models that better match the context. 
Figures \ref{workshop} (a) \& (b) show the 3D generation of "golden white flower" as a lotus flower, given that no explanation is provided by GPT-4. 
However, in the story \textit{Narcissus and the Narcissus Flower}, the ``golden white flower'' represents the narcissus flower. After GPT-4 processes the story and provides a description and explanation, the generation model can produce a result that better aligns with the context of the narrative.  We also found some words that TripoAI previously couldn’t understand, but they were now correctly generated after being explained. As shown in Figures \ref{workshop}(c) \& (d), TripoAI initially failed to understand the meaning of the word ``tuba,'' resulting in the generation of a cylindrical shape. However, after GPT-4 explained, TripoAI successfully generated a tuba.
\paragraph{Design Requirements.} According to the results above, we derived that \textbf{D3:} The main objects should be described and explained in the context of the story before generating.
\subsection{Plausibility of the Generated Content}
We conducted experiments to explore how AI-based reviews and human reviews can be effectively combined to ensure the plausibility of the model.
\paragraph{Apparatus.} We conduct an AI-based review by using CLIP \cite{radford2021learningtransferablevisualmodels}.  Our test set was collected from the generated cases from the experimental group on the second day of the workshop, a total of 183 text-model pairs.
\paragraph{Procedure.} 
 To determine whether a model is plausible, we first asked I1 -- I5 manually to tag the plausibility of the 183 generated models. I1 -- I5 voted on its plausibility. A model is considered plausible if at least three out of five participants ($N>=3$) deem it plausible. Then, we asked I6 to input the text of extracted main characters and objects, along with their corresponding front views of the models, into the CLIP model to calculate CLIP scores \cite{hessel2022clipscorereferencefreeevaluationmetric}. Finally, we evaluated the CLIP score threshold that distinguishes plausible and suspicious models. Let $S$ be the CLIP score of a model, and $c$ be the threshold. Our evaluation starts with $c = 0.9$, decrementing $c$ by 0.1 each time. For each threshold $c$, we calculate the proportion of plausible models among all models with $S>c$.  
 \paragraph{Result} The proportion of plausible models among all 183 models with $S>c$ is shown in Table \ref{tab:reasonable_models}.  Based on the result, we decided that all models with
$S<0.7$ need to be manually reviewed.
\begin{table}[h]
\centering
    \caption{Proportion of Reasonable Models among All Models with $S>c$ for Different Values of \( c \).}
    \label{tab:reasonable_models}
    \begin{tabular}{|c|c|c|c|c|}
        \hline
        $c$ & 0.9 & 0.8 & 0.7 & 0.6 \\ \hline
        \multirow{1}{*}{Proportion of} & \multicolumn{1}{c|}{ } & \multicolumn{1}{c|}{ } & \multicolumn{1}{c|}{ } & \multicolumn{1}{c|}{ } \\ 
        reasonable models & 100\% & 100\% & 100\% & 96\% \\ \hline 
    \end{tabular}
    \vspace{1em}
\end{table}
\begin{figure*}[htbp]
  \centering
  \includegraphics[width=0.98\linewidth]{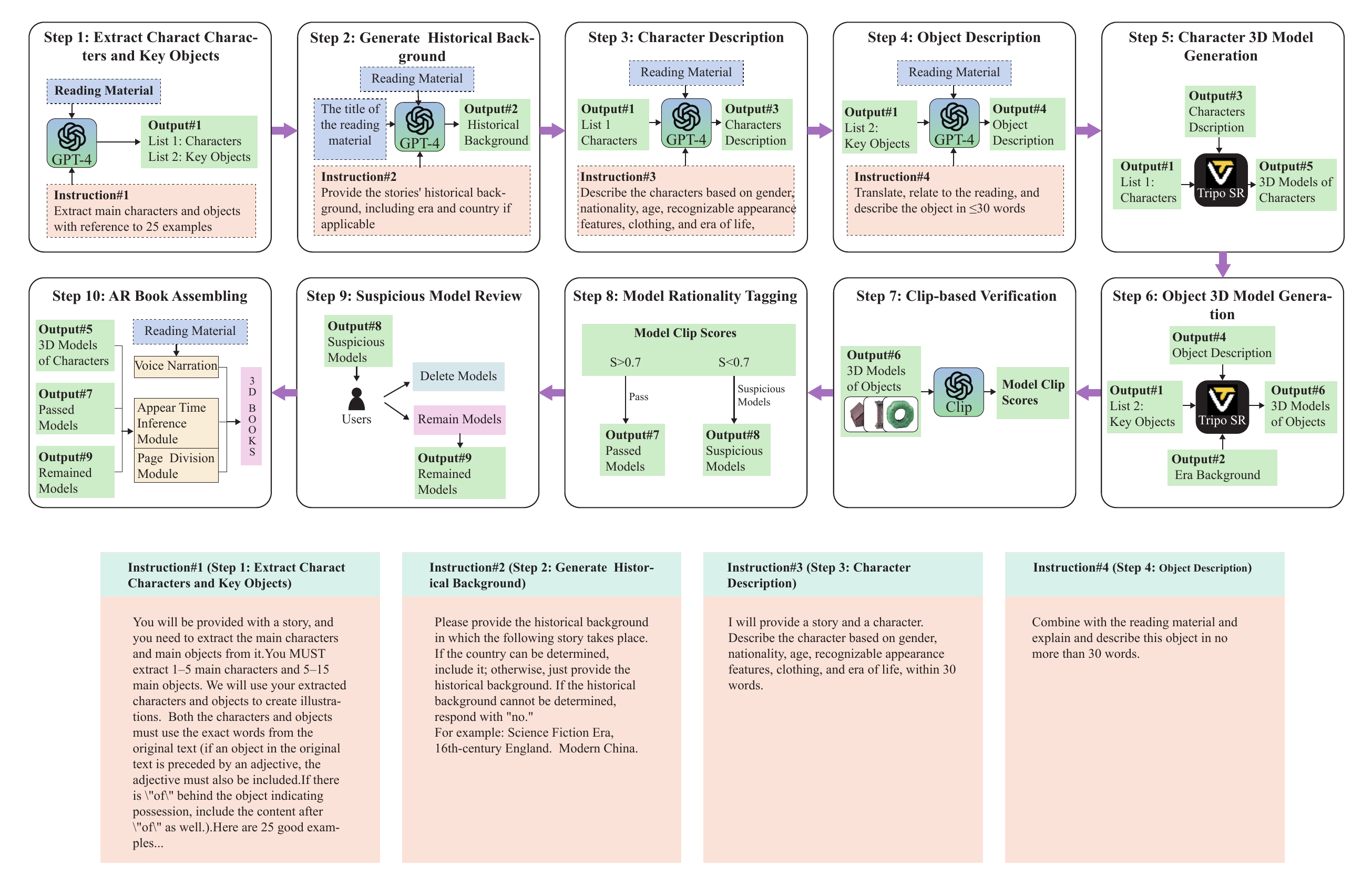}
  \caption{Ten steps of AR book generation pipeline and (the bottom row) the detailed instructions in Steps 1 to 4.}
  \label{fig:pipeline}
  \vspace{-3mm}
\end{figure*}

\paragraph{Design Requirement.} \textbf{D4:} A combination of CLIP detection and human review should be adopted to ensure the plausibility of generated models. Models with a CLIP score below 0.7 are considered suspicious and require manual review.

\section{The Metabook system}
We consider the guidance of four design dimensions according to the workshop findings. We designed and implemented a tool for end-to-end creating 3D books, namely Metabook.
\subsection{Story-to-3D Book Pipeline}
We introduced the “story-to-3D book” pipeline as shown in Figure~\ref{fig:pipeline}. Our pipeline integrates GPT-4 for text processing and TripoSR \cite{tochilkin2024triposrfast3dobject} for model generation. Additionally, we use CLIP to detect model plausibility and manually review suspicious models to ensure the plausibility of the generated models. As shown in Figure~\ref{fig:pipeline}, the story-to-3D book pipeline operates through ten steps without being trained on specific datasets. 
Step 1 extracts the main characters, guided by Instruction\#1 (\textbf{D1}, see Section \ref{sec:workshop}). Step 2 infers the historical background of the story, guided by Instruction\#2.  Since the background of the time and the place could influence the style of objects in the scene, this contextual information enables more precise object descriptions. Step 3 delineates the main characters according to gender, nationality, age, recognizable appearance features, clothing, and the era of life (\textbf{D2}), guided by Instruction\#3. Step 4 elucidates and delineates the main objects (\textbf{D3}), guided by Instruction\#4. Steps 5 and 6 pertain to the creation of 3D models.  Steps 7, 8, and 9 enhance model dependability by the integration of CLIP detection and human evaluation (\textbf{D4}).  Step 10 consolidates all components into the finished 3D book.

\textit{Step 1: Extract characters and main objects.} GPT-4 parses the story text following Instruction\#1. Thus, it extracts the main objects and main characters into a structured output format. As a result, the list of main objects and main characters is the output.

\textit{Step 2: Infer the historical background of the story}
A story often implies or explicitly mentions its historical background. For example, “Louis XVI arrived at the Palace of Versailles in a carriage...” From this, we can infer that the story takes place in 18th-century France. If we input “carriage” directly for 3D generation, there is a risk of generating a modern carriage or a Chinese-style carriage, which would lead to errors. Therefore, in Step 2, we use Instruction\#2, the title of the story, and the story itself as inputs, allowing GPT-4 to help infer the historical context of the story. The purpose of this approach is to ensure that the generated objects better align with the historical and cultural context. 

\textit{Step 3: Character description.} We input the story text and the main characters list, and instruct GPT-4 to describe these characters based on gender, nationality, age, recognizable appearance features, clothing, and the era of life.

\textit{Step 4: Describe and explain the main objects.}  In Step 4, we input the story text and prompt GPT-4 to generate object descriptions and explanations based on the story. The purpose of this approach is to enable GPT-4 to infer object characteristics from the context, ensuring that the generated objects align with the narrative. 

\textit{Steps 5--6: Generation of 3D models.}  We generate 3D models from text by calling the TripoAI text-to-3D API. Regarding characters, our input includes the character’s name from output\#1 and the description of the character from output\#3. As for main objects, our input includes the object’s name from output\#1, the historical background from output\#2, and the context-based description from output\#4.

\textit{Steps 7–9: Ensure the reliability of generated models.}  We have implemented a two-round review process to ensure the models are reasonable. In the first round of review, we use the CLIP model for evaluation. By calculating the CLIP score of the models, we categorize the models into reasonable and suspicious models.  We denote the frontal view image of the generated model as $I_f$.  $O$ represents the corresponding extracted main objects.  The formula for calculating the CLIP score $S$ is as $S=CLIP\left(I_f, O\right)$. 
According to the workshop assessment results, when $S<0.7$, the model is classified as suspicious, and it will be sent for manual review in the second round. 
All suspicious models will be displayed on the front-end interface during the manual review stage. If the uploader elects to remove a model, it will be excluded from the resulting AR book.

\textit{Step 10: 3D book assembling}
To enhance the immersive experience, we use voice narration to tell the story. We utilize Microsoft Azure to convert the story text into speech. The process of assembling the generated 3D models into a 3D book requires the handling of two modules: the page division module and the pop-up time inference module.  Due to the limited space on the 3D book page, we need to control the number of models on each page to be between 4 and 6 (except for the last page, which may have fewer than 4). Therefore, page division needs to be redefined. First, we search for the positions of all extracted keywords (including characters and objects) in the text. Initially, we perform a preliminary page division based on 4 models per page. Then, we consider how to balance the word count on each page. After the preliminary division, we denote the word count of the i-th page as $W_i$. $P_{i,last}$ represents how many words are before the word of the last model on page $i$, and $P_{i,first}$ represents the position of the first model on page $i$. Starting from the first page, when the following condition is satisfied: $W_i-W_{i+1}>P_{i,last}-P_{i,first}$. 
The first model on page i+1 will be moved to page i. The above condition will continue to be checked until it is no longer satisfied or the number of models on page i reaches 6.

Traditional 3D books tend to load models simultaneously for users. However, users may experience reading difficulties caused by receiving excessive visual information at the same time.   Therefore, we gradually display 3D models based on the text sequence. We propose a method based on the speech rate of narration to infer the pop-up time of models. The formula for the pop-up time $T$ is as follows: $T=\left\lceil N_{K} / r \times 5\right\rceil$, where $T$ represents the pop-up time of the 3D model, $ N_{K }$ represents how many words are before the extracted word K, and $r$ represents the speech rate, meaning the number of words read every 5 seconds. Each model will only pop up the first time it is mentioned, ensuring that the same character or object does not appear in different forms.
\subsection{Implementation}
Our system has three components: 1) Smartphone interface: for user upload functionality. The front end is created using Unity and compiled as an APK, which is compatible with any Android device. 2) MR headset (Quest 3): for users to acquire and view 3D literature. In developing the Quest 3 APK, we utilized the see-through video approach for mixed reality, allowing users to see the physical world around them while interacting with the virtual instances. 3) The PC, equipped with an NVIDIA GeForce RTX 3090, functions as a small GPU server, managing the whole process of converting the `story-to-3D book,' facilitating communication between the MR headset and smartphone, and performing data storage, together with the data storage capability.

\subsection{Interaction walkthrough}
\begin{figure*}
  \centering
  \includegraphics[width=\linewidth]{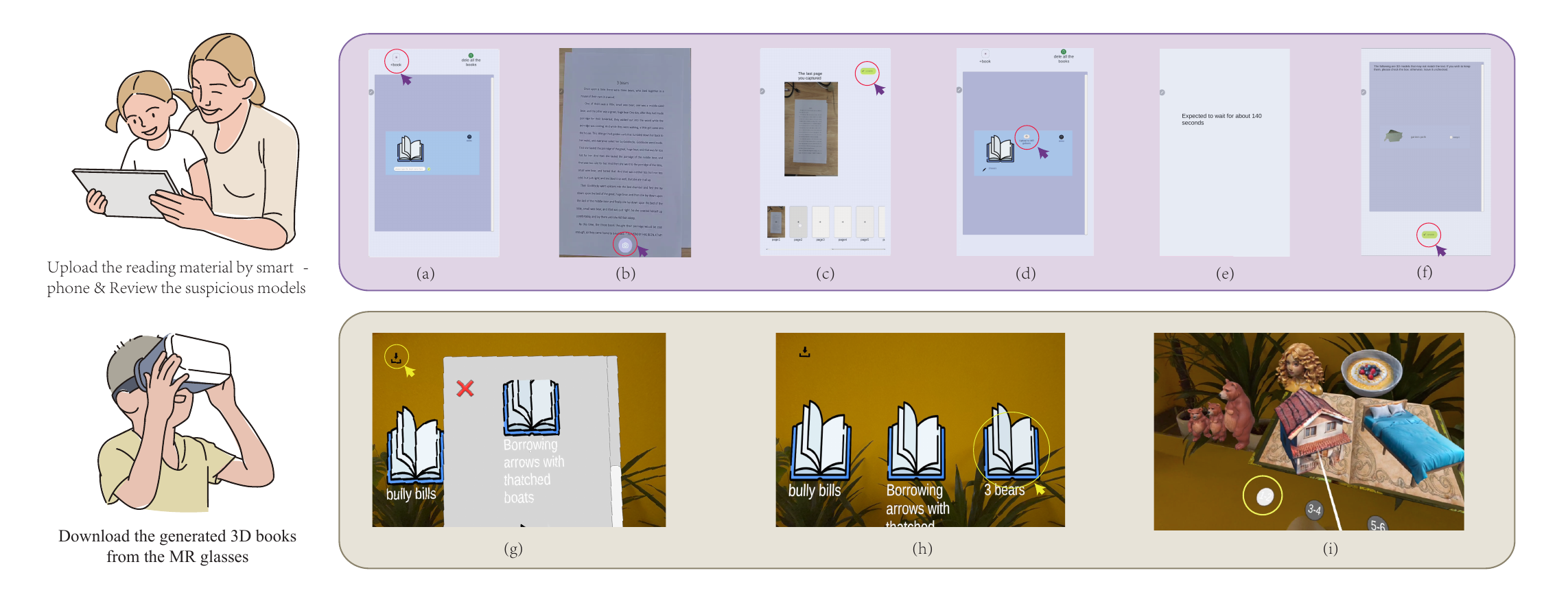}
  \caption{Metaboook's Interaction walkthrough: (a)--(f) book generation on smartphones; (g)--(i) 3D book experiences on MR headsets.}
  \label{fig:UI}
  \vspace{-2mm}
\end{figure*}

Figure~\ref{fig:UI} illustrates the three-stage process that a user follows to produce a 3D book. Imagine a mother called Eliza who intends to produce a 3D book for her kid to read as a bedtime story this evening. She opts for Metabook since it allows anyone without 3D modelling or programming expertise to produce 3D books within a reasonable timeframe. 

\textbf{Upload the reading material by smartphone}. 
Eliza first flipped to the reading material her child was going to read today and clicked the ``+Book'' button (Figure~\ref{fig:UI} (a)) to create a new 3D book. She entered the title of the story, ``Goldilocks and the Three Bears.'' Then, she clicked on the button that looks like an open book to enter the photo-taking interface and took pictures of the content (Figure~\ref{fig:UI} (b)). Afterward, she clicked the ``complete'' button in the upper-right corner to finish taking the photos (Figure~\ref{fig:UI} (c)). Next, she clicked the ``upload to MR glasses'' button (Figure~\ref{fig:UI} (d)) to start creating the 3D textbook. An AR pop-up on the screen with an estimated waiting time of approximately 140 seconds (Figure~\ref{fig:UI} (e)). Eliza took advantage of these two-plus minutes to wash her face.

\textbf{Review the suspicious models }
About three minutes later, Eliza returned to the phone app, and by then, the generation process was complete. She found that the model corresponding to the word ``garden path'' was tagged as a suspicious model. Eliza thinks this model is unreasonable, so she didn't choose ``keep it'' and directly clicked ``complete'' (Figure~\ref{fig:UI} (f)).

\textbf{Download the generated 3D from the MR glasses}
Then Eliza went to her child's room; she instructed her child to download the 3D book named ``Goldilocks and the Three Bears'' to their MR glasses. The child clicked the download button in the upper-left corner to view the available AR books (Figure~\ref{fig:UI} (g)), then selected the book title to begin the download (Figure~\ref{fig:UI} (h)). To start experiencing 3D AR book, the child clicked the page number to view the corresponding page in the 3D book (Figure~\ref{fig:UI} (i)). At that moment, the 3D book would start narrating the content of the page, and the model would appear step-by-step as the narration. 

\section{Study 1: Immersive Book Generation}
\begin{table}
\centering
  \caption{Participants' educational backgrounds and their familiarity with 3D book creation.}
  \label{tab:T1}
  \resizebox{0.5\textwidth}{!}{ 
  \begin{tabular}{ccc}
   \toprule
      \shortstack{Participant \\ ID} & \shortstack{Educational \\ Background} &\shortstack{Familiarity with\\ 3D Books Creation}\\
    \midrule
    \ P1&Master & 1 \\
    \ P2&Bachelor & 1 \\
    \ P3&High School& 1 \\
    \ P4&High School & 1 \\
    \ P5&Associate Degree & 1 \\
    \ P6 &Bachelor &1 \\
    \ P7 &Bachelor &2 \\
    \ P8 &Master &1 \\
    \ P9 &Master &2 \\
    \ P10 &Bachelor &1 \\
    \ P11 &Bachelor &1 \\
  \bottomrule
\end{tabular}
}
  \vspace{-4mm}
\end{table}

The first user evaluation aims to assess the Metabook's user satisfaction and the ease of creating 3D books. We describe the study design and summarize evaluation results in the following paragraphs.

\subsection{Study Design}
\paragraph{Participant}
We recruited 11 adult participants (5 males and 6 females; age range: 23 -- 57 years old; mean: 34.18 years old) through social media. They assume the role of parents who produce immersive books for their children. Table \ref{tab:T1} depicts their educational background as well as prior experience of creating immersive books, following a 5-point scale: 1 denotes “no experience at all” and 5
signifies “highly proficient.” The study was approved
by the university's Institutional Review Board (IRB).

\paragraph{Apparatus and Task Design}
The experiment was conducted in a classroom. We employed an Android smartphone, namely HONOR 80 (Display: 6.67-inch OLED, 1080×2400 resolution; CPU: Qualcomm Snapdragon 782G; RAM: 12GB; Camera: 160MP main + 8MP ultra-wide + 2MP macro), and the participants can upload stories to the smartphone. Then, we provided them with MR headsets, specifically the Meta Quest 3, to experience the generated 3D books in AR of video see-through.

In this task, participants are required to use Metabook to create a 3D book from the story they had prepared. First, they uploaded the story text via an Android phone. Then, they waited for the 3D book to be created. 
Upon receiving the app message indicating the readiness of the 3D book, participants were instructed to return to the app for a manual review.
First, they were instructed to put on MR glasses. Next, they downloaded and experienced the entire 3D book they had created. Finally, participants were asked to evaluate the book creation experience by completing questionnaires, including the system's usability, the satisfaction level of the 3D book quality, and the satisfaction level of generation time.
\begin{figure}[h]
  \centering
  \includegraphics[width=\linewidth]{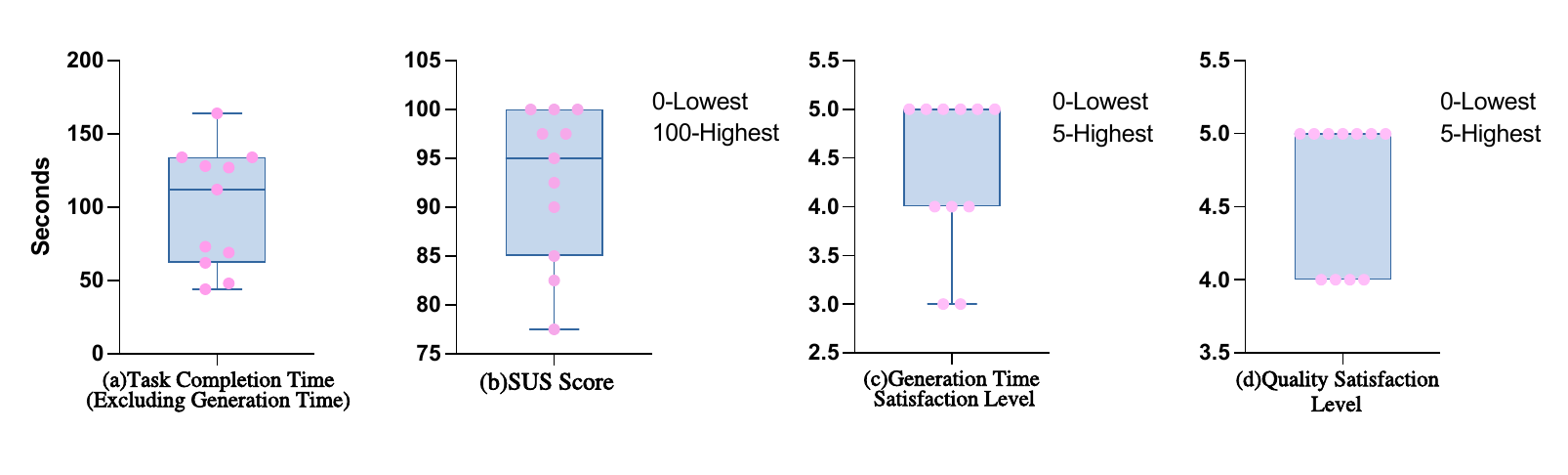}
  \caption{Results of Study 1. The box plots of Task completion time, SUS score,Generation time satisfaction Level, and Quality satisfaction level. The pink dots represent the result values for each participant.}
  \label{fig:sus}
  \vspace{-3mm}
\end{figure}

\paragraph{Procedure} 
Before the experiment began, we instructed participants to independently create story texts intended for conversion into 3D books, which they were to carry throughout the user research. The books possessed by these users included both Chinese and English titles (see Table \ref{tab:t2} for precise titles and lengths).
Before commencing the research, participants had to fill out a background questionnaire that included personal details such as gender, education level, and experience with AR book authoring. Upon entering the classroom, participants first had a 5-minute practical training session to familiarise themselves with the user interface and comprehend the experimental task sequence.
Thereafter, the experimenter instructed individuals to begin the task. Throughout the experiment, the researcher positioned themself at the rear of the classroom, observing participant activity while also documenting the user upload process and task completion time using screen recording. The server's back end will automatically log the time required to produce the 3D book.

\subsection{Results}
Figure~\ref{fig:sus} presents the user experience SUS score on a 5-point scale (1 represents very dissatisfied, 5 represents very satisfied), and the metrics include task completion time (excluding generation time), user satisfaction with the generated 3D book, user satisfaction with the generation waiting time, and user satisfaction with the generated 3D book quality. Table \ref{tab:t2} depicts the names of the generated 3D books, the story length, and the waiting time for each generated 3D book. 

\begin{table}
  \caption{Results of study 1: the 3D book generated by participants and the generation time.}
  \label{tab:t2}
   \resizebox{\columnwidth}{!}{
  \begin{tabular}{ccccc}
   \toprule
           \shortstack{Participant \\ ID}&\shortstack{Article \\ Name}& \shortstack{Word\\ Count} & \shortstack{Generation \\ Time} & \shortstack{Number of \\Models} \\
    \midrule
    \ P1&Egre &361 & 86s & 6 \\
    \ P2&\raggedright Returning the Jade to Kingdom Zhao&805 & 111s & 6 \\
    \ P3&Bird Paradise &1015 & 119s & 7 \\
    \ P4&Watching the Tide &473 & 108s & 6 \\
    \ P5&John Hawkwood &24& 164s & 9 \\
    \ P6&Bully Bill &1348& 174s & 13 \\
    \ P7&Captain Fantastic &889 & 135s & 9 \\
    \ P8&Learning About No &610 & 130s & 8 \\
    \ P9&Mrs.Richards &252 & 143s & 8 \\
    \ P10&The Giraffe &1346 & 163s & 9 \\
    \ P11&\raggedright Borrowing Arrows with Thatched Boats & 1362 & 202s & 15 \\
  \bottomrule
\end{tabular}
\vspace{-2mm}
}
\end{table}
\paragraph{Results of User Experience}
Our system makes it easy for users without 3D book creation experience to create 3D books, from scratch to augmented reality. The average score of SUS is 92.5 (Figure \ref{fig:sus}(b)). Scores above 90 are considered of ``Truly superior products \cite{Asusevaluate}.'' For individuals with no prior experience in creating 3D books, the process typically requires assistance from people with programming and 3D modeling backgrounds. Our system automates the assembly of 3D books and generates 3D models directly from the story text, enabling inexperienced users to create 3D books without relying on professional support. The applied SUS scale has both negative and positive perceptions, where 1 indicates ``strongly disagree'' and 5 indicates ``strongly agree.''. Regarding the negative rating, 
all participants expressed the perception that they would need assistance from a technical expert to use this system, which is equivalent to a 2 out of 5 score . In contrast, 100\% of participants rated ``I felt very confident using the system'', which is a score of 5 on the positive rating. In other words, all participants, after a 5-minute system training, were able to complete the 3D book creation task with great confidence. 

\paragraph{Task Completion Time}
All users completed the upload within 3 minutes (Figure \ref{fig:sus}(a)), significantly saving creators' time compared to traditional 3D book-making methods. Traditional 3D book creation requires a significant amount of time spent on searching for and editing 3D models. Metabook reduces both manual effort and time investment by leveraging an LLM to process story text and using a 3D generation model to generate 3D models.

Our system is user-friendly for people with diverse educational backgrounds. We encode their educational background into degree levels according to the EQF Level \cite{wikipediaEuropeanQualifications}: high school corresponds to level 3, associate degree to level 5, bachelor's degree to level 6, and master's and MPhil to level 7. We calculate the Spearman correlation coefficient $r$ and p-value to examine the relationship between the SUS score and educational background; The result shows that there is no significant correlation between the SUS score and education level (r= 0.07, p= 0.84). This means that the level of education does not significantly affect the user's experience with the system.

\paragraph{Satisfaction with the Generation Time}
All participants rated their satisfaction with the generation time above 3 (where 1 represents very dissatisfied and 5 represents very satisfied), indicating that all participants were satisfied with the generation time (Figure \ref{fig:sus}(c)). In our study, the generation time for the 11 books ranged from a minimum of 86 seconds to a maximum of 202 seconds. We noticed that P1, who had the shortest generation waiting time, gave the lowest satisfaction score of 3 among the 11 participants. On the other hand, P11, who had the longest waiting time, gave the highest satisfaction score of 5. Through the experimenter's notes, we found that P1 was pacing while waiting, while P11 was playing on her phone. Further calculation of the Spearman correlation coefficient r between generation time and satisfaction reveals that there is no significant correlation between generation time and the satisfaction level of generation time (r= 0.13, p= 0.70).  This means that under conditions where users are only waiting for a few minutes, their satisfaction with the waiting time is more likely to be influenced by subjective factors.

\begin{figure}[h]
  \centering
  \includegraphics[width=\linewidth]{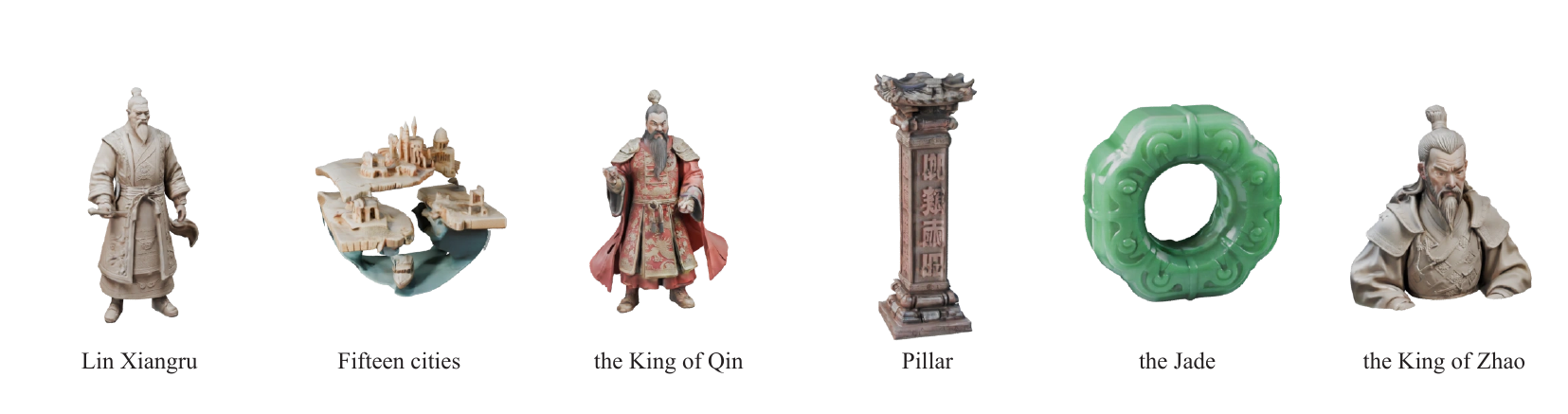}
  \caption{Examples of 3D Models from \textit{Returning the Jade to Kingdom Zhao.}}
  \label{fig:jade}
  \vspace{-2mm}
\end{figure}

\paragraph{Satisfaction with the Generated 3D Book}
All participants rated their satisfaction with the generation time. None of them rated less than 4, with an average satisfaction of 4.64 (where 1 represents very dissatisfied and 5 represents very satisfied).  Figure \ref{fig:jade} shows the generated result of ``Returning the Jade to Kingdom Zhao,'' an ancient Chinese historical story. We can see that the generated objects, such as ``pillar'' and ``jade,'' are contextualized with the historical setting of the story (the Warring States period in China) and its narrative context. For example, the decorations on the pillar reflect traditional Chinese imperial architecture, aligning well with the story’s setting in the Qin royal palace.  The generated characters also align well with their portrayals in the story. For instance, the King of Qin is dressed in red, wearing the most luxurious garments. In the story, he is depicted as the highest-ranking and most fond of extravagance.

\section{Study 2: Impacts of 3D Books on Children's Learning}
The second user evaluation aims to investigate the impact of the created 3D book, in contrast to conventional paper-based reading, on children's learning interest, retention of the story, comprehension, and cognitive load. Our user study follows a within-subjects design, which gathered quantitative and qualitative findings and generated insights.

\subsection{Participants}
We recruited another 22 children, denoted by A1 to A22 (10 boys and 12 girls; age range: 10 -- 13 years old; mean: 11.1 years old), through teachers and parents. Their participation was entirely voluntary. The participants' ages fulfill the minimum age requirement officially set by Meta Quest 3 (i.e., 10-year-olds). None of them had any vision or hearing impairments. The research was done with their guardians' presence, and participants could quit at any point. Our work was authorized and administered by the Research Ethics Board of our university.

\subsection{Task Design}
The participants were told to complete two tasks: paper-based reading (Task 1) and a 3D book experience (Task 2). Two stories of similar difficulty and text length were chosen from a Chinese textbook of the higher grade: "Borrowing Arrows with Straw Boats" and "Wu Song Slays the Tiger." To maintain consistency in our evaluation, our two comprehension quizzes have similar difficulty and none of the participants had read either of the two stories before. 
Regarding the learner's understandings, both quizzes have a maximum score of 10, consisting entirely of single-choice questions. Scoring is based on the correctness of the selected answers. Next, participants are required to recall and summarize both stories in writing, i.e., memory retention. After the summarization is completed, three teachers separately evaluate the summary result based on two aspects: accuracy and completeness. Each teacher scores the summary on a scale from 0 (worst) to 10 (best), and we compute the average scores. Then, participants are required to complete the NASA-TLX \cite{HART1988139} questionnaire. Finally, the participants were asked to complete two Smileyometers and two Again-Again tables \cite{read2002endurability,yung2018printy3d} to measure their interest in these two modes. The participants were instructed to focus on the two learning modes, instead of the reading material itself. 

\subsubsection{Task 1: Paper-based reading}
The participants need to complete reading ``Wu Song Fights the Tiger'' in 8 minutes. After reading the paper book, we collect their reading materials and ask them to complete a reading comprehension quiz within 12 minutes. 

\subsubsection{Task 2: 3D book experience}
\begin{figure}[hpt!]
  \centering
  \includegraphics[width=\linewidth]{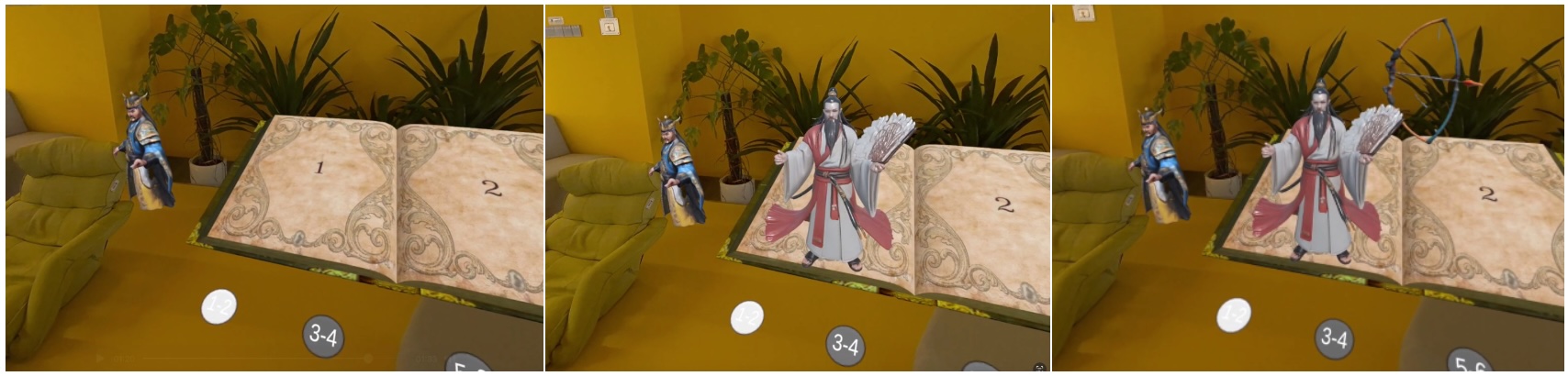}
  \caption{First-person view of \textit{Borrowing Arrows with Straw Boats } on a MR headset.}
  \label{fig:mr_pig}
  \vspace{-2mm}
\end{figure}

As shown in Figure \ref{fig:mr_pig}, the participants need to complete experiencing the 3D book ``Borrowing Arrows with Straw Boats'' in an MR headset within 8 minutes. After completing the MR experiences, we help them take off the MR glasses and ask them to complete the comprehension quiz within 12 minutes. 

\subsection{Procedure} 
Before the experiment began, the participants spent 15 minutes receiving a tutorial on how to use the system and the Meta Quest 3, as well as being informed about the experimental procedure. We aided the participants in wearing and calibrating the Quest 3, ensuring their comfort and the clarity of their perception of both the virtual and real environments. Then, the participants were required to complete Task 1 and Task 2, followed by a semi-structured interview.  Finally, we invited six teachers (TS1 to TS6) to experience the 3D book and conducted semi-structured interviews with them.

\subsection{Result}
In this section, we first perform the Shapiro-Wilk test to determine whether the data follow a normal distribution. We conduct a paired sample t-test for normally distributed data, while for non-normally distributed data, we perform the Wilcoxon signed-rank test. Then, we discuss the impact of 3D books on children's interest of two learning modes, memory of stories, comprehension of stories, and cognitive load. Finally, we summarize and discuss teachers' perspectives on using generated 3D books for children's education.

\begin{figure}[hpt!]
  \centering
  \includegraphics[width=1.1\linewidth]{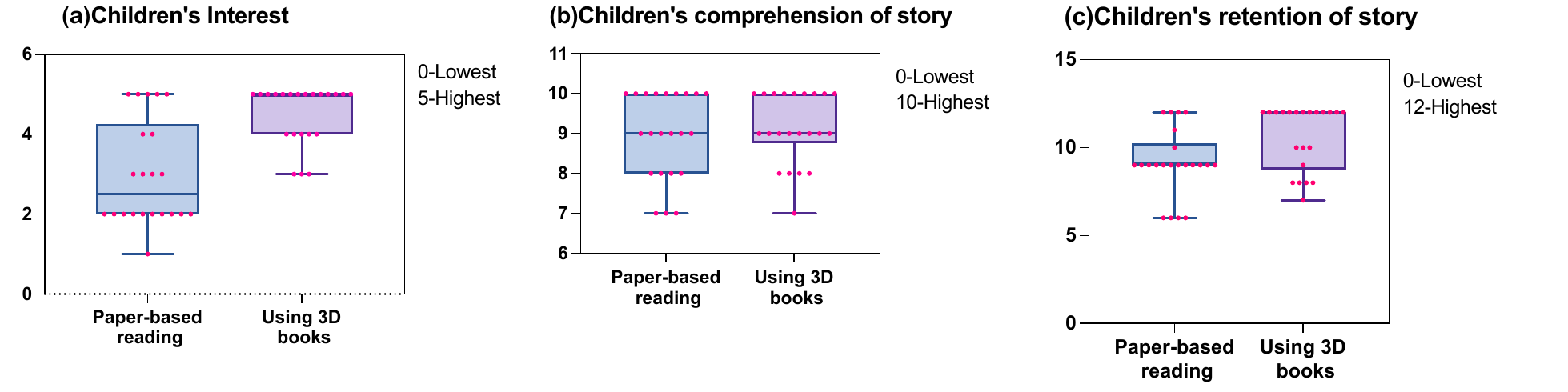}
  \caption{Results of Study 2. The pink
dots represent the result values for each participant. }
  \label{fig:3imgs}
\end{figure}

\paragraph{Children's interest}

Overall, children's  interest significantly increased through the use of Metabook (Figure \ref{fig:3imgs}(a)). In the Smileyometer ratings for the level of interest, statistical significance is observed ($Z=-3.4078, p <0.001$). Group A gave scores of 3 out of 5 for the paper book reading mode and 4.5 out of 5 for the 3D book  mode.  13 children mentioned in interviews that they liked the “3D illustrations” in the book. Child A16 said, “I like the pictures in this magic book because they are three-dimensional, vivid, and interesting. They make me feel relaxed, and I can clearly see the scenes in the story.”  

A notable difference was observed in the children's willingness to learn the next story ($Z=-3.5162, p <0.001$). Their willingness to learn another story also increased significantly when they used a 3D book. The participants scored 2 out of 3 for their willingness to learn another story with the paper medium, while they gave a strong score of 3 out of 3 for their willingness to learn another story using augmented reality. When using traditional paper books, only 27.28\% of children were willing to learn another story, whereas 100\% of children were willing to learn another story when using the 3D book. This evidence reflects that children with Metabook are more willing to continue learning.

\paragraph{Children's memory of story content}
Students' retention of the story content significantly improved by using a 3D book (Figure \ref{fig:3imgs}(c)). In the recall and retell sections, students scored 9.1 out of 12 and 10.6 out of 12, respectively, after using traditional paper books and 3D books, showing an improvement of 1.5 points ($Z= -2.495, p < 0.05$). This indicates that using Metabook enhanced children’s overall memory of the story, making it easier for them to recall the complete story content.

\paragraph{Comprehension of stories}
Children demonstrated comparable levels of story comprehension across both modes ($Z=-0.6816, p > 0.05$). Figure \ref{fig:3imgs}(b) shows the average scores of children's comprehension quizzes (a 10-point scale). The scores for traditional paper books and 3D books are 8.95 and 9.13, respectively. A possible reason is that comprehension of the story content is influenced by high-level skills (inferences and the ability to interpret and integrate information) \cite{STOLE2020103861}, and using the generated 3D book does not help improve these high-level skills. The number of people scoring full marks (10) is the same for both modes (Figure \ref{fig:3imgs}(b)). Remarkably, the gap in reader performance for Metabook is narrower than the traditional counterpart, e.g., the number of people scoring the lowest (7) has decreased. This result gives clues that Metabook, featuring interactivity and immersiveness, may help those with less learning performance. 

\paragraph{Cognitive load}
\begin{figure}[hpt!]
  \centering
  \includegraphics[width=\linewidth]{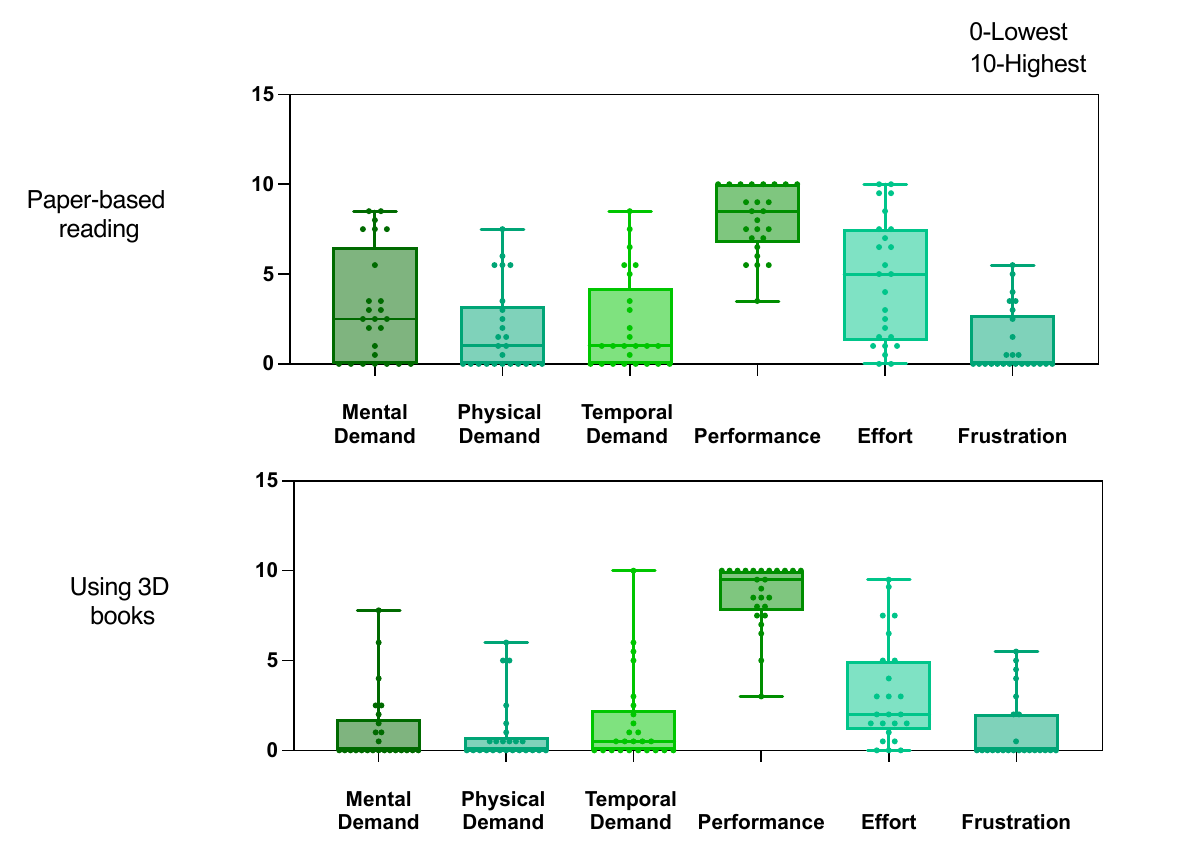}
  \caption{Examples of 3D Models from \textit{Returning the Jade to Kingdom Zhao.}}
  \label{fig:nasa}
  \vspace{-3mm}
\end{figure}
Figure \ref {fig:nasa} shows the user workload between the two conditions, and using the 3D book can reduce the cognitive load of children, in terms of Mental Demand ($Z=-2.33, p<0.05$) and Effort ($Z=-2.31, p<0.05$). There is an absence of statistical significance in other metrics. The results imply that when story content of similar difficulty and length, using the 3D book makes children feel subjectively more relaxed and experience less mental burden.

\paragraph{Teachers' perception of Metabook}
In the interview, six teachers of linguistic background, with teaching experiences (in years): $\bar{M}=10.67; SD=8.73; Min=3; Max=21$. 
They shared their insights on the use of Metabook in fostering children's learning motivations. Overall, the teachers expressed a positive outlook on Metabook's ability to ignite interest in learning among young learners. Several teachers highlighted its unique approach to storytelling, which combines verbal and visual elements to captivate children's imaginations. TS2 explained, ``\textit{Young children primarily use visual thinking. Metabook builds a bridge between verbal and visual thinking, making learning more engaging for children.}''  TS5 noted, ``\textit{Children are easily attracted by the 3D visuals}'' TS3 also stated, ``\textit{This storytelling method can spark children's curiosity about upcoming plot developments.}'' 
Beyond sparking interest in learning, some educators recognized Metabook's potential to enhance classroom dynamics. TS3 mentioned,  ``\textit{Metabook potentially demonstrates the ability to invigorate the classroom atmosphere and re-engage students during quieter moments, noting that it effectively captures their attention.}'' TS4 saw Metabook as a valuable tool for inspiring creative writing, suggesting that when children input their compositions into Metabook, they become more immersed in the process, which can stimulate their creativity. 

However, three teachers (TS1, TS4, TS6) also voiced concerns regarding the use of Metabook for learning. TS1 expressed apprehension that the captivating 3D illustrations might distract children, potentially hindering their focus on learning. TS6 further commented, ``\textit{Children might develop a dependence on these electronic devices, so I would not recommend excessive use.}'' These insights reflect a balanced perspective on the integration of Metabook into educational settings, highlighting both its benefits and potential challenges.

\section{Concluding Remarks and Discussion}
Metabook is a comprehensive pipeline that allows novice users to create 3D books in AR . Responding to our research questions, our studies validated the system’s usefulness. A subsequent comparative study revealed that 3D book can enhance children’s interest, improve memory retention, and reduce cognitive load, though it does not significantly improve comprehension of stories. Existing prior works have demonstrated diverse aspects of children's learning in emerging multimedia, while our results contribute to this ongoing conversation in a comprehensive view, with the emphasis on leveraging generative AI and multi-modal learning. We highlight our findings in the following discussion.

\textbf{(A) 3D Book's Impact on Learning. }
The results from Study 2 demonstrate that Metabook can significantly enhance children's interest. This is in line with previous observations on children’s interests \cite{rambli2013fun,dibrova2016ar,chemerys2022combined,lukcyhasnita2024development}.  With our generated 3D book involving 3D illustrations and voice narration, young readers feel more interested, active and enthusiastic about learning. 
Regarding story retention, our findings suggest that 3D books help children perform better in retelling the story, which aligns with earlier studies by Matthew et al. \cite{matthew1997comparison} and Danaei et al. \cite{danaei2020comparing}. Since children rely on both textual and visual information to make meaning, 3D books provide multimedia features that aim to enhance children’s attention by offering key imagery elements, which may help them remember more episodes. 
With respect to story comprehension, our findings revealed no significant difference between 3D book and traditional paper book reading, which is inconsistent with prior research \cite{danaei2019influence,danaei2020comparing,csimcsek2023effects}.  One possible explanation is that prior studies used AR books that included animation, while our AR 3D books deliberately used static models to minimize cognitive load for children \cite{buchner2022impact}. Comprehension primarily measures the child's understanding of the characters’ thoughts and emotions, as well as the main idea and message of the story. Animations can better show the characters' facial expressions and manners than static content \cite{danaei2020comparing}. This is consistent with the findings reported by Takacs and Bus \cite{takacs2016benefits}, who found that animated books led to greater story comprehension in children than static versions.

\textbf{(B) Cognitive Load. } 
Although previous studies have shown that using AR in education carries the risk of increasing cognitive load, potentially distracting children and affecting learning outcomes, such increases in cognitive load are often associated with information overload \cite{buchner2022impact}. Prior research has demonstrated that inappropriate design elements — such as including details unrelated to the narrative — may interrupt information processing by imposing additional cognitive load \cite{danaei2020comparing,kao2016effects,10674259,csimcsek2024examining}. In our study, we provide evidence that AR 3D books, featuring computationally selected content and reasonable design dimensions, can effectively reduce the amount of information presented simultaneously, i.e., by displaying 3D models step by step. 
Supported by the results of the final evaluation, we meticulously leveraged the combination of visual and auditory modalities to reduce the burden on any single channel. This enabled us to use the benefits of AR to augment learning engagement and enhance the learning experience, without imposing additional cognitive burdens on youngsters. It is essential for MR and educational system developers to regulate cognitive strain while proficiently creating MR educational solutions. Methods include segmenting information, using multimodal presentations, and utilizing visual cues to direct attention.

\textbf{(C) Potential Applications of 3D Books. }
3D books are appropriate for instructors in contexts where students experience boredom or disengagement, and when retention of information is crucial. For instance, the retention of terminology, the recollection of historical events, and the acquisition of scientific formulae. Nonetheless, in disciplines necessitating comprehensive analysis and comprehension, reliance just on 3D books does not yield substantial advancements and may require instructor elucidation and direction to augment students' higher-order skills~\cite{STOLE2020103861}. Simultaneously, when students exhibit less enthusiasm for studying, the utilisation of 3D books might enhance their inclination to persist in their studies \cite{dong2018research}. 3D books may be used to facilitate the learning of complex material, hence mitigating the adverse psychological states pupils may encounter when confronted with challenges.  Simultaneously, when using 3D books, it is crucial to regulate the duration of student engagement to prevent overdependence on the digital gadget. 

\textbf{(D) Limitations and Future Work. }
While our system garnered favourable feedback and our research yielded significant insights for educators and MR developers, several limitations must be recognized. First, our user study was conducted in a nation with a homogeneous ethnic composition, resulting in all participants sharing the same ethnic background and language proficiency. We consider our work a foundational step that can be expanded in future studies to include individuals from varied ethnic backgrounds. Furthermore, it is crucial to note that the relatively small sample size (Workshop: N = 6; User Evaluation: N = 11 in Study 1 and N = 22 in Study 2) in this study constrains the generalizability of our results. Although this study illustrates the system's utility, it is restricted by the experiment's duration and participant count. The longest narrative utilized to create a 3D book in this study comprised 1,362 characters. Future research should strive to incorporate lengthier narratives and a larger participant pool.








\bibliographystyle{plain}
\bibliography{template}

\end{document}